\RequirePackage[utf8]{inputenc}
\RequirePackage[T1]{fontenc}
\documentclass[reprint,aps,pra,superscriptaddress,floatfix]{revtex4-1}
\pdfoutput=1

\usepackage{mathtools}
\usepackage{braket}
\usepackage{bbold}

\usepackage[normalem]{ulem}

\usepackage[single]{acro}

\DeclareAcronym{CFI}{
	short = CFI,
	long = classical Fisher information,
}
\DeclareAcronym{QFI}{
	short = QFI,
	long = quantum Fisher information,
}
\DeclareAcronym{CRB}{
	short = CRB,
	long = Cramér-Rao bound,
}
\DeclareAcronym{QCRB}{
	short = QCRB,
	long = quantum Cramér-Rao bound,
}
\DeclareAcronym{CSL}{
	short = CSL,
	long = continuous spontaneous localisation,
}
\DeclareAcronym{POVM}{
	short = POVM,
	long = positive-operator valued measure,
}
\DeclareAcronym{SLD}{
	short = SLD,
	long = symmetric logarithmic derivative,
}

\usepackage{siunitx}
\DeclareSIUnit\rad{rad}

\DeclareMathOperator{\diag}{diag}

\newcommand\LambdaR{\lambda}
\newcommand\LambdaSQL{\Lambda_{0}}
\newcommand\omegat{\tau}
\newcommand\thermvar{\kappa_{\text{th}}}

\newcommand\rS{r_\mathrm{s}}
\newcommand\rC{r_\mathrm{C}}
\newcommand\lambdaCSL{\lambda^{\mathrm{CSL}}}

\newcommand\commutator[2]{[#1,#2]}
\newcommand\estimator[1]{\widetilde{#1}}

\newcommand\identity{\mathbb{1}}
\newcommand\op[1]{\hat{#1}}
\NewDocumentCommand\Trace{m}{
	\operatorname{Tr} \left( #1 \right)
}
\newcommand\unitful[1]{\MakeUppercase{\mathcal{#1}}}
\newcommand\unitfulop[1]{\unitful{\op{#1}}}

\usepackage[colorlinks=true,citecolor=blue,linkcolor=red,urlcolor=blue]{hyperref}

\begin{document}
\title{Quantum enhanced estimation of diffusion}

\author{Dominic Branford}
\affiliation{Department of Physics, University of Warwick, Coventry, CV4 7AL, United Kingdom}

\author{Christos N. Gagatsos}
\affiliation{College of Optical Sciences, University of Arizona, 1630 E. University Blvd., Tucson, Arizona 85719, United States of America}
\affiliation{Department of Physics, University of Warwick, Coventry, CV4 7AL, United Kingdom}

\author{Jai Grover}
\author{Alexander J. Hickey}
\affiliation{ESA---Advanced Concepts Team, European Space Research Technology Centre (ESTEC), Keplerlaan 1, Postbus 299, NL-2200AG Noordwijk, The Netherlands}

\author{Animesh Datta}
\affiliation{Department of Physics, University of Warwick, Coventry, CV4 7AL, United Kingdom}

\date{2 September 2019}

\begin{abstract}
	Momentum diffusion is a possible mechanism for driving macroscopic quantum systems towards classical behaviour.
	Experimental tests of this hypothesis rely on a precise estimation of the strength of this diffusion.
	We show that quantum-mechanical squeezing offers significant improvements, including when measuring position.
	For instance, with \SI{10}{\dB} of mechanical squeezing, experiments would require a tenth of proposed free-fall times.
	Momentum measurement is better by an additional factor of three, while another quadrature is close to optimal.
	These have particular implications for the space-based MAQRO proposal---where it could rule out the spontaneous collapse theory due to Ghirardi, Rimini, and Weber---as well as terrestrial optomechanical sensing.
\end{abstract}

\maketitle

\section{Introduction}

Finding a unified description of microscopic and macroscopic systems remains an enduring quest of fundamental physics.
One class of proposed solutions are collapse models~\cite{bassi_dynamical_2003,bassi_models_2013,bassi_uniqueness_2013,bassi_gravitational_2017} which span \ac{CSL}~\cite{pearle_combining_1989,ghirardi_markov_1990,toros_colored_2017}, Karolyhazy~\cite{karolyhazy_gravitation_1966}, Diósi-Penrose~\cite{diosi_universal_1987,diosi_models_1989,penrose_gravitys_1996,bahrami_role_2014}, and quantum gravity~\cite{ellis_quantum_1989}; as well as collisional decoherence~\cite{gallis_environmental_1990}.
In the non-relativistic regime, they posit spatial decoherence due to diffusion in momentum.
The outcome is a description of the evolution in terms of a phase-space density distribution obeying a Fokker-Planck diffusion equation~\cite{ghirardi_unified_1986}.
Experimental advances have now made the testing of this proposition a realistic prospect.

Mechanical systems have been used to bound the strength of such diffusive effects.
Examples include gravitational-wave detectors~\cite{carlesso_experimental_2016}, the LISA pathfinder experiment~\cite{carlesso_experimental_2016,helou_lisa_2017,carlesso_non-interferometric_2018}, ultracold cantilevers~\cite{vinante_improved_2017}, and trapped ions~\cite{li_detecting_2017}.
Proposals for future experiments which could probe collapse models and further study macroscopic quantum states include the generation of macroscopic superpositions~\cite{romero-isart_large_2011,romero-isart_quantum_2011,scala_matter-wave_2013,wan_free_2016,bose_spin_2017,weaver_phonon_2018} and the space-based MAQRO mission~\cite{kaltenbaek_macroscopic_2012,kaltenbaek_macroscopic_2016} which formed a key focus of a recent ESA feasibility study~\cite{european_space_agency_cdf_2018}.
	
\begin{figure}[htbp]
	\centering
	\includegraphics{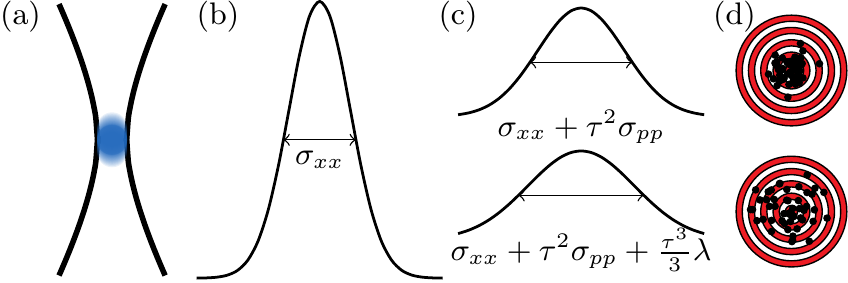}
\caption{A pictorial representation of the measurement of wavepacket expansion which forms part of the MAQRO proposals~\cite{kaltenbaek_macroscopic_2012,kaltenbaek_macroscopic_2016}.
	(a) A particle is initially trapped, (b) then released, (c) the free particle wavefunction expands, more rapidly with a localisation term, (d) localisation rate can be inferred through position measurements.
Expansion as depicted in two spatial detections is for illustrative purposes, we only analyse one independent spatial dimension.}
	\label{fig:maqro_schematic}
\end{figure}

One simple experiment---which forms a part of the MAQRO mission~\cite{kaltenbaek_macroscopic_2012,kaltenbaek_macroscopic_2016}---to test collapse models is to let free particles evolve and measure the expanding width of the wavepacket.
Once all classical noise sources have been ruled out, any excess wavepacket width must be attributed to momentum diffusion associated with collapse models.
MAQRO aims to utilise ultracold nanoparticles and exploit the nano-gravity of space to observe free-fall over \SI{100}{\s}---enabling more precise sensing of momentum diffusion---as represented in Fig.~\ref{fig:maqro_schematic}.

Quantum techniques such as squeezing allow for more precise estimation~\cite{toth_quantum_2014,demkowicz-dobrzanski_quantum_2015}.
Optical squeezing has been identified as valuable to fundamental physics, with squeezing-enhanced interferometry~\cite{caves_quantum-mechanical_1981} set to enhance laser-interferometric gravitational-wave detectors~\cite{the_ligo_scientific_collaboration_gravitational_2011,grote_first_2013,the_ligo_scientific_collaboration_enhanced_2013} and \SI{15}{\dB} squeezing of optical vacuum reported~\cite{vahlbruch_detection_2016}.
It has also found application in photonic-force microscopy~\cite{taylor_fundamental_2013,taylor_subdiffraction-limited_2014}, while microwave squeezing is being used in the search for axion dark matter~\cite{malnou_squeezed_2019}.

In this article, we show that quantum squeezing of the mechanical degree of freedom enables a more precise estimation of the strength of momentum diffusion. 
This enhancement is attainable with the currently proposed scheme of measuring the position of a particle.
Quantum squeezing of mechanical degrees of freedom is beginning to be explored in thermal states~\cite{pontin_squeezing_2014,rashid_experimental_2016}.
We conclude that squeezing can be used to achieve the same precision with reduced free-fall time or centre of mass cooling.
This reduction could be ten-fold for a squeezing of \SI{10}{\dB}.
Thus, squeezing can compensate for reduced free-fall times, identified as one of the challenges for MAQRO~\cite{kaltenbaek_macroscopic_2012,kaltenbaek_macroscopic_2016} in a recent ESA CDF study~\cite{european_space_agency_cdf_2018}.
We further show that a momentum measurement is thrice as precise as that of position, while measurement of a more general quadrature is close to optimal.
We briefly discuss the potential of the heterodyne and phonon counting measurements.

While our results will be presented in the context of collapse models, observing similar momentum diffusion processes could aid detection of certain dark-matter candidates~\cite{riedel_direct_2013,riedel_decoherence_2015,riedel_decoherence_2017}.
Since excess heating of wavepackets is also a consequence of momentum diffusion~\cite{collett_wavefunction_2003,diosi_testing_2015,li_detecting_2017}, our results imply a quantum enhanced estimation of heating.
Finally, the ubiquitous phenomena of Brownian motion is also caused by diffusion.
Our results can thus be applied in this very general scenario, as well as in particle tracking used to study biological systems~\cite{ghislain_scanning-force_1993,pralle_local_1998}.

Before presenting our results, we note some recent works that have theoretically considered continuously monitoring a thermal state~\cite{genoni_unravelling_2016} or squeezing a specific optomechanical coupling~\cite{mcmillen_quantum-limited_2017}, with the latter providing no attainable advantage from squeezing when measuring the optical subsystem.
Previous works in quantum metrology have analysed quantum-limited estimation of related noise parameters including loss~\cite{monras_optimal_2007,adesso_optimal_2009}, diffusion in phase shifts~\cite{knysh_estimation_2013,vidrighin_joint_2014} and displacements~\cite{tsang_quantum_2019}, and classical stochastic processes~\cite{ng_spectrum_2016}.

\section{Background}

A particle of mass \( m \) in a harmonic potential has Hamiltonian
\(
	\unitfulop{H}
	=
	\unitfulop{p}^2/2m
	+
	m \omega^2 \unitfulop{x}^2/2.
\)
Dimensionless position and momentum operators are
\( \op{x} =  \sqrt{m\omega}\unitfulop{x}/\sqrt{\hbar} \)
and
\( \op{p} = \unitfulop{p} / \sqrt{\hbar m\omega} \) 
whose commutators are given by the matrix \( i\Omega \) where
\begin{equation}
	\Omega = -i \begin{pmatrix} \commutator{\op{x}}{\op{x}} & \commutator{\op{x}}{\op{p}} \\ \commutator{\op{p}}{\op{x}} & \commutator{\op{p}}{\op{p}}  \end{pmatrix} = \begin{pmatrix} 0 & 1 \\ -1 & 0 \end{pmatrix}.
	\label{eq:commutator}
\end{equation}

Quantum states of such a particle have a phase-space representation in terms of the Wigner function of an operator defined as~\cite[Chap.~1]{ferraro_gaussian_2005}
\begin{equation}
	W_{\rho}(x,p) = \frac{2}{\pi} \Trace{\rho \op{D}\left(\frac{x+ip}{\sqrt{2}}\right) \op{\Pi} \op{D}^{\dagger}\left(\frac{x+ip}{\sqrt{2}}\right) },
	\label{eq:wigner}
\end{equation}
where \( \op{D}(\alpha) = e^{\alpha\op{a}^{\dagger}-\alpha^*\op{a}} \) and \( \op{\Pi} = e^{i\pi \op{a}^{\dagger}\op{a}} \).
Gaussian states are those whose Wigner function is Gaussian and so determined by the averages---displacement vector \( \vec{d} \)---and covariances---covariance matrix \( \sigma \)---of the position and momentum operators.
Examples include thermal, coherent, and squeezed states.
A thermal state has covariance matrix
	\( \sigma = \thermvar \identity \),
with \( \sigma = \identity \) corresponding to the ground state.

We focus on the simplest setup to study momentum diffusion, that of a free particle as in Fig.~\ref{fig:maqro_schematic}.
Initially the particle is trapped in a harmonic potential with frequency \( \omega \) and cooled.
Cooling of nano-particles has been reported to the order of \num{100} phonons~\cite{windey_cavity-based_2019,delic_cavity_2019} with theory anticipating cooling much closer to the ground state~\cite{romero-isart_quantum_2012,gonzalez-ballestero_theory_2019}.
After cooling the trapping potential is turned off.
The particle then evolves freely under the Hamiltonian
\( \unitfulop{H} = \unitfulop{P}^2/2m \) with Lindblad term \( \Lambda \commutator{\unitfulop{X}}{\commutator{\unitfulop{X}}{\rho}} \), whose strength \( \Lambda \) is our parameter of interest.
The master equation for momentum diffusion for this system---in terms of the dimensionless position and momentum operators---is
\begin{equation}
	\frac{\partial \rho}{\partial \omegat} = - \frac{i}{2} \commutator{\op{p}^2}{\rho} - \frac{1}{4} \LambdaR \commutator{\op{x}}{\commutator{\op{x}}{\rho}},
	\label{eq:unitless_master}
\end{equation}
where \( \omegat = \omega t \) and \( \LambdaR = \Lambda / \LambdaSQL \) are dimensionless parameters, and \( \LambdaSQL = m\omega^2/(4\hbar) \). 
Being quadratic the master equation Eq.~\eqref{eq:unitless_master} evolves Gaussian states to Gaussian states~\cite{carmichael_statistical_1999,nicacio_phase_2010,serafini_quantum_2017}.

Eq.~\eqref{eq:unitless_master} can then be transformed to a Fokker-Planck equation~\cite{barnett_methods_2002,nicacio_phase_2010}, in this case yielding
\begin{equation}
	\frac{\partial}{\partial \omegat} W(x,p,\omegat) = \left[ - p \frac{\partial}{\partial x} + \frac{1}{4} \LambdaR \frac{\partial^2}{\partial p^2} \right] W(x,p,\omegat),
	\label{eq:fokker}
\end{equation}
which for Gaussian \( W \) can be mapped to the equations of motion of form~\cite{carmichael_statistical_1999,serafini_quantum_2017}
\begin{equation}
\begin{aligned}
	\frac{\partial\vec{\mu}}{\partial \tau} &=  A \vec{\mu}, &
	\frac{\partial\sigma}{\partial \tau} &= A \sigma + \sigma A^T + D,
	\label{eq:eom_gaussian}
\end{aligned}
\end{equation}
where \( \vec{\mu} \) and \( \sigma \) are the Gaussian's moments.
For an initial Gaussian state with moments \( \vec{d} \) and \( \sigma \) the evolved moments under Eq.~\eqref{eq:fokker} become
\begin{align}
	\vec{d}(\omegat) &=  \begin{pmatrix} 1 & \omegat \\ 0 & 1 \end{pmatrix} \vec{d},
	\label{eq:disp}\\
	\sigma(\omegat) &= \begin{pmatrix} 1 & \omegat \\ 0 & 1 \end{pmatrix} \sigma \begin{pmatrix} 1 & 0 \\ \omegat & 1 \end{pmatrix} + \LambdaR \begin{pmatrix} \omegat^3/3 & \omegat^2/2 \\ \omegat^2/2 & \omegat \end{pmatrix}.
	\label{eq:cov}
\end{align}

Our results apply to estimation of diffusion in any scenario governed by Eq.~\eqref{eq:unitless_master} for all values of \(\LambdaR \) and \( \omegat \).
To estimate the strength of the momentum diffusion \( \Lambda \), we begin with a single-mode Gaussian state.
Such a state can be described as a thermal state \( \thermvar\identity \) with a squeezing \( r \geq 0 \) of the quadrature \( \op{x}\sin\phi+\op{p}\cos\phi \) giving an initial covariance matrix
\begin{equation}
	\sigma = \thermvar
	\begin{pmatrix}
		\cosh 2r + \sinh 2r \cos 2\phi & \sinh 2r \sin 2\phi \\
		\sinh 2r \sin 2\phi & \cosh 2r - \sinh 2r \cos 2\phi 
	\end{pmatrix},
\label{eq:covariance_initial}
\end{equation}
with arbitrary displacements.
The displacements do not begin with any parameter-dependence and do not gain any through the evolution given by Eq.~\eqref{eq:disp} and so their derivative with respect to the parameter satisfies \( \partial_\Lambda \vec{d} = 0 \).
We will consider tuning \( \phi \) to maximise the precision for given thermal variance and squeezing magnitudes, with \( \phi=0 \) and \( \phi=\pi/2 \) corresponding to momentum and position squeezing respectively. 

We will highlight special cases for \( \LambdaR \ll 1 \) and \( \omegat \gg 1 \), which is the regime for MAQRO~\cite{kaltenbaek_macroscopic_2012,kaltenbaek_macroscopic_2016} as in Table~\ref{tab:maqro_values}; and \( \thermvar \sim 1 \) which is around the MAQRO regime.

\begin{table}[htbp]
	\centering
	\begin{ruledtabular}
	\begin{tabular}{cc}
		Localisation rate \( \Lambda \) & \SIrange{1e10}{1e20}{\m^{-2}\s^{-1}} \\
		Free-fall time \( t \) & \SI{100}{\s} \\
		Mechanical frequency \( \omega \) & \SI{1e5}{\rad\per\s} \\
		Mass \( m \) & \SIrange{1e8}{1e10}{\amu} \\
		Thermal occupation number \( n_{\text{th}} \) & \num{0.3} \\
		Thermal variance (\( \thermvar  = 2n_{\text{th}}+1 \)) & \num{1.6} \\
		Limiting localisation\footnotemark[1] (\( \LambdaSQL = \frac{m\omega^2}{4\hbar} \)) & \SI{1.6e26}{\m^{-2}\s^{-1}} \\
		Experiment timescale (\( \omegat = \omega t \)) & \num{6.3e7}
	\end{tabular}
	\end{ruledtabular}
	\footnotetext[1]{Using \( m = \SI{1e8}{\amu} \)}
	\caption{Parameter values based on \citet{kaltenbaek_macroscopic_2016}, primarily Table~1 therein.}
	\label{tab:maqro_values}
\end{table}

An estimator is required to estimate an unknown parameter from observed data.
If limited to statistical noise the precision of the value produced by the estimator can be taken from the variance of that estimator.
The \ac{CRB} lower bounds the variance of an unbiased estimator as~\cite{kay_fundamentals_1998,helstrom_quantum_1976,holevo_probabilistic_2011,paris_quantum_2009}
\begin{equation}
	(\Delta \estimator{\Lambda})^2 \geq \frac{1}{\nu F(\Lambda)} \geq \frac{1}{\nu H(\Lambda)},
	\label{eq:CRBs}
\end{equation}
where \( \nu \) is the number of repetitions of an experiment, \( \widetilde{\Lambda} \) is an estimator of the parameter \( \Lambda \), and \( F(\Lambda) \) and \( H(\Lambda) \) are respectively the \ac{CFI} and \ac{QFI}.
The \ac{CFI} is a function of the probability distribution~\cite{kay_fundamentals_1998}
\begin{equation}
	F(\Lambda) = \int\! \mathrm{d}\mkern-1mu\vec{x} \, \frac{1}{P(\Pi_{\vec{x}}|\rho_{\Lambda})} \left( \frac{\partial P(\Pi_{\vec{x}}|\rho_{\Lambda})}{\partial \Lambda} \right)^2,
	\label{eq:cfi}
\end{equation}
where the probabilities \( P(\Pi_{\vec{x}}|\rho_{\Lambda}) \) are derived from applying the \ac{POVM} \( \boldsymbol{\Pi} \) to the state \( \rho_{\Lambda} \).
The \ac{QFI} is a function of the state alone~\cite{paris_quantum_2009,toth_quantum_2014,demkowicz-dobrzanski_quantum_2015}
\begin{equation}
	H(\Lambda) = \Trace{\rho_{\Lambda} L_{\Lambda}^2},
	\label{eq:qfi}
\end{equation}
where \( L_{\Lambda} \) is the \ac{SLD} defined by \( L_{\Lambda} \rho_{\Lambda} + \rho_{\Lambda} L_{\Lambda} = 2\partial_{\Lambda} \rho \).

These \ac{CFI} and \ac{QFI} provide the \ac{CRB} and \ac{QCRB}, the first and second inequalities of Eq.~\eqref{eq:CRBs} respectively.
The equalities in Eq.~\eqref{eq:CRBs} are obtained by an optimal measurement, where it exists, and an efficient estimator; we identify such a measurement and the maximum likelihood estimator is asymptotically efficient~\cite{kay_fundamentals_1998}.

For a Gaussian state (where \( \partial_{\Lambda} \vec{d} = 0 \)) the \ac{QFI} can be evaluated explicitly as~\cite{monras_phase_2013,safranek_quantum_2015}
\begin{equation}
	H(\Lambda) = \frac{1}{2} (\partial_{\Lambda}\sigma|(\sigma\otimes\sigma-\Omega\otimes\Omega)^{-1}|\partial_{\Lambda}\sigma) ,
	\label{eq:gaussian_qfi}
\end{equation}
where the inner product is \( (A|B) = \Trace{A^T B} \).

\section{Results}

Using Eqns.~\eqref{eq:cov} and~\eqref{eq:covariance_initial}, the \ac{QCRB} can be calculated through Eq.~\eqref{eq:gaussian_qfi} to be
\begin{equation}
	(\Delta\estimator{\Lambda})^2 \geq
	\frac{
	\LambdaSQL^2\left[
		\left(\thermvar^2 + \omegat \thermvar \LambdaR Z
		+ \frac{\omegat^4}{12} \LambdaR^2 \right)^2
		- 1
		\right]
	}{
		\frac{\omegat^4}{12}\left( 1 - \thermvar^2 
		+ \omegat \thermvar \LambdaR Z
		+ \frac{\omegat^4}{12} \LambdaR^2 \right)
		+ \frac{\omegat^2}{2}\thermvar^2  Z^2 
	},
\label{eq:qcrb_squeeze}
\end{equation}
where
\(
Z = \left(1+\omegat^2/3\right)\cosh 2r + \left[ \left(1-\omegat^2/3\right) \cos 2\phi + \omegat\sin 2\phi \right]\sinh 2r.
\)
The bound in Eq.~\eqref{eq:qcrb_squeeze} behaves as \( (\Delta\estimator{\Lambda})^2 \gtrsim \Lambda^2 \) to leading order in \( \Lambda \).

The \ac{QCRB} in Eq.~\eqref{eq:qcrb_squeeze} is minimised by squeezing or anti-squeezing (squeezing the orthogonal quadrature) with squeezing angle (See App.~\ref{app:optimal_squeezing})
\begin{equation}
	\phi  =\arctan \left( \frac{ -3+\omegat^2 - \sqrt{9+3\omegat^2+\omegat^4} }{ 3\omegat } \right),
	\label{eq:qcrb_optimal_squeeze}
\end{equation}
which tends to \( 0 \) for \( \omegat \gg 1 \), corresponding to squeezing of position or momentum.
When squeezing at this angle in the regime of \( \omegat \gg 1 \), with \( \thermvar = 1 \), the \ac{QCRB} simplifies to
\[
	(\Delta\estimator{\Lambda})^2 \gtrsim
	\LambdaSQL^2 
	\frac{8\LambdaR\left( e^{-2r} + \frac{\omegat}{4}\LambdaR \right) \left( 1 + \frac{\omegat^3}{6} e^{-2r} \LambdaR + \frac{\omegat^4}{24} \LambdaR^2 \right)}{
		\frac{2\omegat^3}{3} e^{-4r} + \frac{\omegat^4}{3}e^{-2r}\LambdaR + \frac{\omegat^5}{12}\LambdaR^2
	},
\]
with the squeezing \( r \) not necessarily positive as anti-squeezing may be preferrable (see App.~\ref{app:optimal_squeezing}).

Measurement of the particle's position is a special case of homodyne detection which involves measuring a linear combination of the position and momentum quadratures~\cite{adesso_continuous_2014,serafini_quantum_2017}.
Heterodyne allows for the simultaneous measurement of position and momentum, but with added noise~\cite{shapiro_phase_1984,leonhardt_measuring_1995}.
The \ac{QCRB} can be reached through projection onto eigenstates of the \ac{SLD}~\cite{braunstein_statistical_1994} which, for a Gaussian system, entails performing some squeezing and displacement followed by measurement of Fock states~\cite{monras_optimal_2007,monras_phase_2013,serafini_quantum_2017}.
This additional squeezing is a resource applied to the system after the evolution as part of the measurement and does not improve the precision as an initial squeezing can.
Further, in a mechanical system this involves measuring the number of phonons which remains experimentally demanding~\cite{cohen_phonon_2015,hong_hanbury_2017}.
In the following, we calculate the performance of all these measurements for estimating \( \Lambda \).

Homodyne detection at an angle \( \theta \) measures the quadrature
\(
	\op{q}_{\theta} = \op{x}\cos\theta+\op{p}\sin\theta
\).
When performed on a Gaussian state the homodyne statistics are Gaussian~\cite{adesso_continuous_2014} and the moments are the appropriate marginal of the Wigner function.
For a homodyne angle \( \theta \) the variance of the marginal is
	\begin{equation}
\begin{aligned}
	\Sigma &= 
		\thermvar \bigg[
			\left[ (1 + \omegat^2)\cos^2 \theta + \omegat \sin 2\theta + \sin^2 \theta \right]\cosh 2r \\
			&\mkern56mu+ \big\{ 
				\left[ (1-\omegat^2)\cos^2 \theta - \omegat \sin 2\theta - \sin^2 \theta \right]\cos 2\phi \\
				&\mkern112mu+ \left[ 2\omegat \cos^2 \theta + \sin 2\theta \right]\sin 2\phi
			\big\}\sinh 2r \\
		&\mkern56mu+ \LambdaR \left( \frac{\omegat^3}{3}\cos^2 \theta + \frac{\omegat^2}{2}\sin 2\theta + \omegat \sin^2 \theta  \right)
		\bigg],
\end{aligned}
	\label{eq:homodyne_moments}
	\end{equation}
as the Wigner function's mean is parameter-independent so is the marginal's.
The choices \( \theta = 0 \) and \( \theta = \pi/2 \) correspond to measurement of position and momentum respectively.
We will consider the optimisation of \( \theta \), which more generally requires measuring a linear combination of the position and momentum operators.

For a Gaussian probability distribution with a parameter-independent mean, the \ac{CFI} is~\cite[Chap.~3]{kay_fundamentals_1998}
\begin{equation}
	F(\Lambda) = \frac{1}{2} \Trace{\Sigma^{-1} \partial_{\Lambda}\Sigma \Sigma^{-1} \partial_{\Lambda}\Sigma},
	\label{eq:gaussian_cfi}
\end{equation}
where \( \Sigma \) is the variance of the Gaussian distribution.
Using Eqns.~\eqref{eq:homodyne_moments} and~\eqref{eq:gaussian_cfi}, the \ac{CRB} for homodyne along an angle \( \theta \) is
\begin{widetext}
\begin{equation}
\begin{aligned}
	(\Delta\estimator{\Lambda})^2 \geq 
	2\LambdaSQL^2 \Bigg[
		\LambdaR + \thermvar 
		\Bigg(&
			\frac{ \omegat^2 \cos^2\theta + \omegat \sin 2\theta + 1
			}{
				\frac{\omegat^3}{3}\cos^2\theta + \frac{\omegat^2}{2} \sin 2\theta + \omegat\sin^2\theta
			} \cosh 2r \\
			&\mkern64mu-
			\frac{
				\left(\omegat^2 \cos^2\theta + \omegat \sin 2\theta - \cos2\theta \right)\cos 2\phi
				-
				\left( 2\omegat\cos^2 \theta + \sin 2\theta \right)\sin 2\phi
			}{
				\frac{\omegat^3}{3}\cos^2\theta + \frac{\omegat^2}{2} \sin 2\theta + \omegat\sin^2\theta
			} \sinh 2r
		\Bigg)
	\Bigg]^2.
\end{aligned}
\label{eq:crb_homodyne_squeezed}
\end{equation}
To leading order in \( \Lambda \) this is \( (\Delta\estimator{\Lambda})^2 \gtrsim 2\Lambda^2 \) which occurs when the first term in the square dominates, whereas when that can be neglected the bound is a \( \Lambda \)-independent constant.
The bound on estimating the diffusion \( \Lambda \) from position (\( \theta = 0 \)) measurement is
\begin{equation}
	(\Delta\estimator{\Lambda})^2 \geq
	2\LambdaSQL^2
	\left[ \LambdaR + \thermvar \left( 
	\frac{
		\left[ 1 + \omegat^2 \right]\cosh 2r + \left\{ \left[ 1 - \omegat^2 \right]\cos 2\phi + 2\omegat\sin 2\phi \right\}\sinh 2r
	}{\omegat^3 / 3}
	\right) \right]^2,
	\label{eq:crb_position_squeezed}
\end{equation}
\end{widetext}
which behaves as
\begin{equation}
	(\Delta\estimator{\Lambda})^2 \gtrsim 2\LambdaSQL^2 \left[ \LambdaR + \thermvar \frac{\cosh 2r - \sinh 2r \cos 2\phi}{\omegat/3} \right]^2,
	\label{eq:crb_position_omegat}
\end{equation}
for \( \omegat \gg 1 \).
Instead for measuring the momentum (\( \theta = \pi/2 \)) the bound on estimating the diffusion \( \Lambda \) is
\begin{equation}	
	(\Delta\estimator{\Lambda})^2 \geq
	2\LambdaSQL^2
	\left[ \LambdaR + \thermvar 
	\frac{\cosh 2r - \sinh 2r\cos 2\phi}{\omegat}
\right]^2,
	\label{eq:crb_momentum_squeezed}
\end{equation}
which (neglecting squeezing) matches the large \( \omegat \) limit of position measurements when \( \LambdaR \gg \thermvar / \omegat \) and is a factor of 9 better when \( \LambdaR \ll \thermvar / \omegat \).

The optimal input squeezing angle \( \phi \) can in general be found by minimising the coefficient of \( \sinh 2r \) in Eq.~\eqref{eq:crb_homodyne_squeezed} which gives
\begin{equation}
	\phi = - \arctan \left( \frac{1}{\omegat + \tan\theta} \right).
	\label{eq:optimal_hom_squeezing_angle}
\end{equation}
For momentum measurements (\( \theta = \pi/2 \)) this squeezing angle is \( \phi = 0 \) (squeezing of momentum).
While for position measurements (\( \theta = 0 \)) this is \( \phi = - \arctan(1/\omegat) \) tending to \( \phi = -\pi/2 \) for \( \omegat \ll 1 \), and \( \phi = 0 \) for \( \omegat \gg 1 \).

In general the squeezing angle in Eq.~\eqref{eq:optimal_hom_squeezing_angle} produces a precision
\begin{equation}
	(\Delta\estimator{\Lambda})^2 \geq 
	2\LambdaSQL^2 \left[
		\LambdaR + \thermvar e^{-2r} \chi(\omegat,\theta)
\right]^2,
\label{eq:crb_homodyne_optimally_squeezed}
\end{equation}
from which the unsqueezed case (\( r = 0 \)) can also be extracted,
where
\begin{equation}
	\chi(\omegat,\theta) =
	\frac{ \omegat^2 \cos^2\theta + \omegat \sin 2\theta + 1}{ \frac{\omegat^3}{3}\cos^2\theta + \frac{\omegat^2}{2} \sin 2\theta + \omegat\sin^2\theta }.
\end{equation}
One effect of squeezing is equivalent to an effective reduction of \( \thermvar \) by \( e^{-2r} \), unlike reducing the centre-of-mass motion which reaches \( \thermvar = 1 \) at absolute zero this squeezing allows an unlimited reduction in the second term.
For \( \omegat \gg 1 \) (as \( \chi \sim 1/\omegat \)) the same squeezing could instead be considered as an effective increase in \( \tau \) by a factor of \( e^{2r} \) to obtain the same precision from a much shorter free-fall time.

When the quadrature given by Eq.~\eqref{eq:optimal_hom_squeezing_angle} is squeezed the homodyne angle which minimises the bound in Eq.~\eqref{eq:crb_homodyne_optimally_squeezed} is
\begin{equation}
	\theta = - \arctan \left( \frac{3+2\omegat^2+\sqrt{9+3\omegat^2+\omegat^4}}{3\omegat} \right),
	\label{eq:optimal_homodyne}
\end{equation}
which tends to \( \theta \approx -\pi / 2 + 1 / \omegat \) for \( \omegat \gg 1 \).
Measuring the quadrature given by Eq.~\eqref{eq:optimal_homodyne} with squeezing as Eq.~\eqref{eq:optimal_hom_squeezing_angle} gives a precision
\begin{equation}
	(\Delta\estimator{\Lambda})^2 \geq 
	2\LambdaSQL^2 \left[
		\LambdaR + \thermvar e^{-2r} 
		\frac{3+\omegat^2-\sqrt{9+3\omegat^2+\omegat^4}}{\omegat^3/2}
\right]^2.
\label{eq:crb_optimal_homodyne_optimally_squeezed}
\end{equation}

Measuring the quadrature of Eq.~\eqref{eq:optimal_homodyne} does not in general attain the \ac{QCRB}.
When \( \LambdaR \) dominates, the \ac{QCRB} behaves as \( \Lambda^2 \) while any homodyne terms tend to \( 2\Lambda^2 \).
In the \( \tau \gg 1 \) regime, one could improve on the precision by no more than a factor of 2 using heterodyne detection (see App.~\ref{app:heterodyne}).
Fig.~\ref{fig:homodyne_heterodyne_optimality} suggests that heterodyne otherwise shows little promise.

Phonon counting---in combination with displacement and squeezing operations---can in principle attain the \ac{QCRB} for all \( \LambdaR \) and \( \omegat \) as the \ac{SLD} is a quadratic operator in the quadrature operators~\cite{monras_phase_2013,serafini_quantum_2017} and so has eigenstates which are squeezed-displaced Fock states.
The additional squeezing required to attain the \ac{QCRB} is derived in full generality in Appendix~\ref{app:sld_projectors}.
For MAQRO, this squeezing seems nugatory, with \( \SI{79}{\dB} \) required to attain the \ac{QCRB} for \( \Lambda = \SI{1e20}{\m^{-2}\s^{-1}} \) which would only improve precision by a factor of \( \sqrt{2} \), to \( \SI{158}{\dB} \) for \( \Lambda=\SI{1e10}{\m^{-2}\s^{-1}} \),  where the improvement on position measurements would be more pronounced.
In other scenarios, however, this could be worthwhile. 
For \( \omegat \ll 1 \) and \( \LambdaR \omegat^2 \lesssim 1 \) the squeezing needed is only \( e^{2z} \approx 1 + \omegat \approx 1 \), while for \( \omegat \gg 1 \) and \( \LambdaR \omegat \gtrsim 1 \) this goes to \( e^{2z} \approx 2\omegat/\sqrt{3} \).

\section{Discussion}
\begin{figure}[htb]
	\centering
	\includegraphics{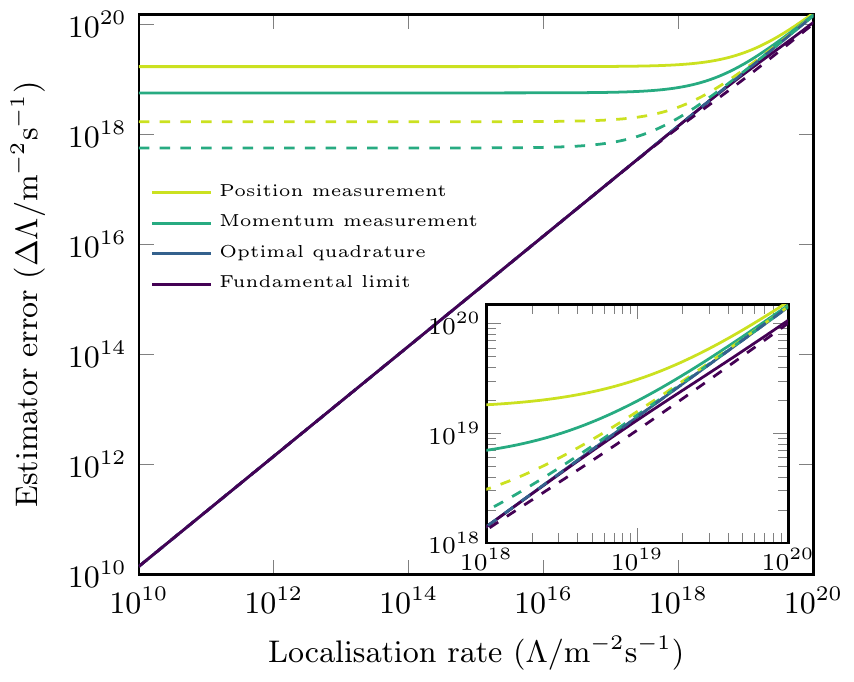}
	\caption{Precision of estimating momentum diffusion from wavepacket expansion for MAQRO parameters (Tab.~\ref{tab:maqro_values}).
	Dashed lines denote a squeezing of \SI{10}{\dB}.
	The optimal homodyne and fundamental limit lines overlap until around \( \Lambda \sim \SI{1e20}{\m^{-2}\s^{-1}} \).
	Three years data collection with \( t = \SI{100}{\s} \) yields \( \nu \sim \num{1e6} \) repetitions.
}
	\label{fig:maqro_precision}
\end{figure}
Fig.~\ref{fig:maqro_precision} shows the potential improvement in precision for estimating diffusion via momentum or homodyne measurements, or through squeezing, for MAQRO parameters as given in Tab.~\ref{tab:maqro_values}.
For reference, position measurement is the present proposal.
We propose squeezing of the momentum quadrature which offers a substantial improvement across much of the pertinent \( \Lambda \) range for both measurement of position and momentum, with \( \SI{10}{\dB} \) enabling an order of magnitude higher resolution of \( \Lambda \).
Measuring the quadrature described by Eq.~\eqref{eq:optimal_homodyne} allows further improvement keeping within a factor of two of the \ac{QCRB} across the whole regime.

Our bounds can be mapped to the wealth of diffusive processes whose parameters enter into the observed diffusion rate \( \Lambda \).
In the case of (mass-proportional) \ac{CSL} the two parameters of interest are \( \lambdaCSL \) and \( \rC \)---the time and length scales in the model.
The observed diffusion rate \( \Lambda \) for a free sphere of mass \( m \) and radius \( \rS \) is---as a function of \( \lambdaCSL \) and \( \rC \)---given by~\cite{collett_wavefunction_2003,kaltenbaek_macroscopic_2016}
\begin{equation}
	\Lambda = \frac{\lambdaCSL}{4\rC^2} \left( \frac{m}{m_0} \right)^2 f\!\left( \frac{\rS}{\rC} \right),
\end{equation}
where \( m_0 \) is a reference (nucleon) mass and 
\(
	f(x) = 
	\frac{6}{x^4} \left[ 1 - \frac{2}{x^2} + \left( 1 + \frac{2}{x^2} \right) e^{-x^2} \right] .
\)
From this bounds on \( \lambdaCSL \) as a function of \( \rC \) can be calculated using
\begin{equation}
	\Delta \lambdaCSL = 4 \rC^2 \left[ \left( \frac{m}{m_0} \right)^2 f\left( \frac{\rS}{\rC} \right) \right]^{-1} \Delta \Lambda.
\end{equation}

To describe the minimal discernable \( \lambdaCSL \) for measurement of a mechanical quadrature we take the limit of the single-shot \ac{CRB}
\( \lambdaCSL_0 = \lim\limits_{\lambdaCSL \to 0} \Delta \lambdaCSL \).
Allowing for \( \nu \) independent repetitions the uncertainty can be reduced to \( \Delta \lambdaCSL \approx \sqrt{\frac{1}{\nu}} \left( \sqrt{\frac{1}{\nu}} + 1 \right) \lambdaCSL_0 \) at \( \lambdaCSL \approx \lambdaCSL_0 / \sqrt{\nu} \).
To ensure any deviation can be recognised with statistical significance we take the minimum detectable collapse rate \( \lambdaCSL_{\mathrm{min}} \) to be \( \lambdaCSL_{\mathrm{min}} \sim \frac{2}{\sqrt{\nu}} \lambdaCSL_0 \).
Thus, for a quadrature measurement the minimum resolvable \( \lambdaCSL \) we take to be given by \( \lambdaCSL_{\mathrm{min}} = \frac{2}{\sqrt{\nu}} \lim\limits_{\lambdaCSL \to 0} \Delta\lambdaCSL \) in Eq.~\eqref{eq:crb_homodyne_squeezed}.

\begin{figure}[htb]
\centering
\includegraphics{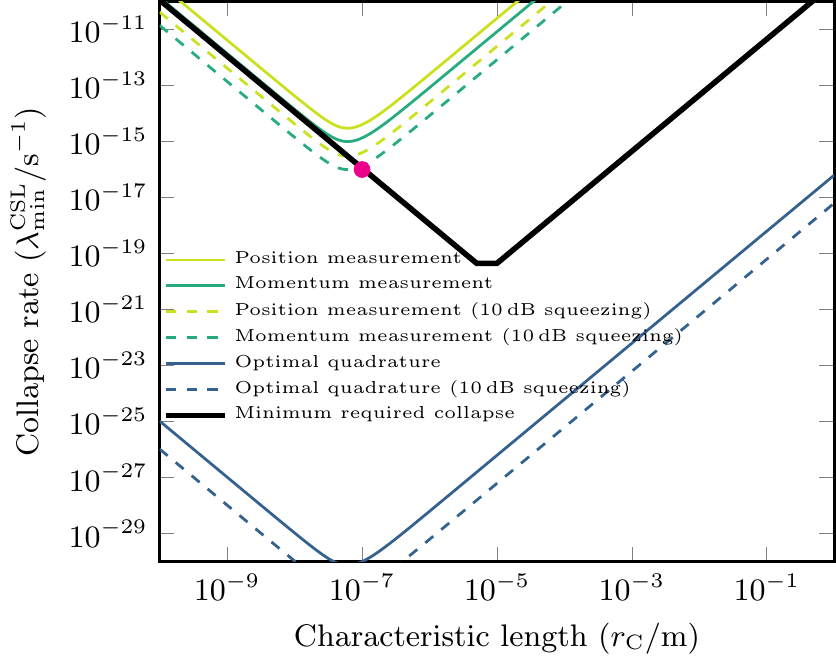}
\caption{
	Minimum detectable collapse rate for three years of observation with a \SI{100}{\nm} radius sphere of mass \SI{5.5e9}{\amu}, with other parameters as Tab.~\ref{tab:maqro_values}.
	The minimum required collapse rate given is based on the criteria of Ref.~\cite{toros_colored_2017} to ensure macroscopic objects rapidly collapse to classical states.
	The magenta dot represents the values originally proposed by \citet{ghirardi_unified_1986}.
}
\label{fig:maqro_csl}
\end{figure}

For MAQRO such bounds can be seen in Fig.~\ref{fig:maqro_csl} for the position, momentum, and optimal quadratures.
For position or momentum measurements with up to \SI{10}{\dB} squeezing the bounds are competive across \SIrange{1e-8}{1e-5}{\m}, below \SI{1e-8}{\m} X-ray emission data begins to provide a tighter bound~\cite{piscicchia_csl_2017} while above \SI{1e-5}{\m} LISA Pathfinder data is tighter~\cite{carlesso_experimental_2016,carlesso_non-interferometric_2018}.
Additional squeezing can of course further reduce the undertainty, with \SI{20}{\dB} of squeezing sufficient to match the theoretical minimum collapse rate to above \SI{1e-7}{\m}.
This would include testing the original parameters suggested by \citet{ghirardi_unified_1986}.

The optimal quadrature identified in Eq.~\eqref{eq:optimal_homodyne} meanwhile could yield a conclusive test of the conventional \ac{CSL} model at a precision of six orders of magnitude more than the theoretical lower bound on \ac{CSL}~\cite{toros_colored_2017}.
Attaining the \ac{QCRB} can offer further improvements, however this would be of little value to MAQRO if the optimal homodyne sensitivity can be reached.

In conclusion, we have shown that squeezing could be used to compensate for reduced free-fall times, an aspect which a recent ESA CDF study~\cite{european_space_agency_cdf_2018} has identified as one of the more demanding of the original proposals~\cite{kaltenbaek_macroscopic_2012,kaltenbaek_macroscopic_2016}.
As---for both Eq.~\eqref{eq:crb_position_omegat} and Eq.~\eqref{eq:crb_momentum_squeezed}---the precision is constant for \( e^{2r}\omegat \) being constant, longer effective free-fall times can be generated through mechanical squeezing.
We have also shown the efficacy of momentum and general quadrature measurements over the proposed position measurement.

\begin{acknowledgments}
We thank Rainer Kaltenbaek, Hendrik Ulbricht, Matteo Carlesso, and Francesco Albarelli for illuminating discussions.
This study has been supported by the European Space Agency's Ariadna scheme (Study Ref.\ 17-1201a), the UK EPSRC (EP/K04057X/2), and the UK National Quantum Technologies Programme (EP/M01326X/1, EP/M013243/1).
D.B.\ has received support for travel and attendance at workshops from QTSpace (COST Action CA15220).
\end{acknowledgments}

\bibliography{maqro}

\appendix

\onecolumngrid

\section*{Appendix}

Appendix~\ref{app:optimal_squeezing} calculates the necessary squeezing angle to maximise precision for the fundamental limit and quadrature measurements.
Appendix~\ref{app:heterodyne} calculates the \ac{CRB} of heterodyne measurements.
Appendix~\ref{app:sld_projectors} derives the necessary squeezing required to then project onto the eigenstates of the \ac{SLD} by phonon counting.
Appendix~\ref{app:relative_precisions} compares performance of the fundamental limit, optimal quadrature, and heterodyne measurements.
Appendix~\ref{app:csl} translates the bounds on the observed diffusion rate \( \Lambda \) to the parameters of \ac{CSL}.

\section{Optimal squeezing}
\label{app:optimal_squeezing}
\subsection{Fundamental limit}
\label{app:qcrb_squeezing}
The \ac{QCRB} is
\begin{equation}
	(\Delta\Lambda)^2 \geq
	B =
	\LambdaSQL^2
	\frac{
		\left(\thermvar^2 + \omegat \thermvar \LambdaR Z + \frac{\omegat^4}{12} \LambdaR^2 \right)^2
		- 1
	}{
		\frac{\omegat^4}{12}\left( \thermvar^2 
		+ \omegat \thermvar \LambdaR Z
		+ \frac{\omegat^4}{12} \LambdaR^2 \right)
		+\frac{\omegat^4}{12}\left( 1-2\thermvar^2 \right) + \frac{1}{2}\thermvar^2 \omegat^2 Z^2 
	},
\label{eq:app:qcrb_squeeze}
\end{equation}
where
\begin{equation}
	Z = \left(1+\frac{\omegat^2}{3}\right)\cosh 2r + \left[ \left(1-\frac{\omegat^2}{3}\right) \cos 2\phi + \omegat\sin 2\phi \right]\sinh 2r.
\end{equation}
Minima with respect to the squeezing angle of the bound in Eq.~\eqref{eq:app:qcrb_squeeze} are either solutions of \( \frac{\partial B}{\partial Z} = 0 \) or \( \frac{\partial Z}{\partial \phi} = 0 \) as
\begin{equation}
	\frac{\partial B}{\partial \phi} = \frac{\partial B}{\partial Z} \frac{\partial Z}{\partial \phi},
\end{equation}
and the second derivative
\begin{equation}
	\frac{\partial^2 B}{\partial \phi^2} = \frac{\partial^2 B}{\partial Z^2} \left(\frac{\partial Z}{\partial \phi}\right)^2
	+ \frac{\partial B}{\partial Z} \frac{\partial^2 Z}{\partial \phi^2},
\end{equation}
distinguishes minima and maxima.
The stationary points of \( B(Z) \) are
\begin{equation}
	Z_{\pm} = \frac{
		144(1-\thermvar^4) + 24\LambdaR^2\omegat^4(1-2\thermvar^2) + \LambdaR^4\omegat^8 \pm \left|12(1-\thermvar^2)+\LambdaR^2\omegat^4\right| \sqrt{[12(1+\thermvar^2)+\LambdaR^2\omegat^4]^2-48\LambdaR^2\thermvar^2\omegat^4} 
	}{
		288 \LambdaR \thermvar^3 \tau
	},
\end{equation}
where the negative root is not possible with \( r > 0 \) and for the positive root
\(
	\frac{\partial^2 B}{\partial Z^2} < 0,
\)
means that the minimum of \( B \) is found for \( \frac{\partial Z}{\partial \phi} = 0 \).
The stationary points of \( Z(\phi) \) are
\begin{equation}
	\phi_{\pm} = \arctan \left( \frac{-3+\omegat^2 \pm \sqrt{9+3\omegat^2+\omegat^4} }{ 3\omegat } \right),
\end{equation}
where we have
\begin{equation}
	(\phi_+ - \phi_-) \mod \pi = \frac{\pi}{2}
\end{equation}
as \( \tan (\phi_+) \tan (\phi_-) = -1 \).
Hence we recognise that squeezing the quadrature \( \op{x}_{\phi_+} \) is equivalent to anti-squeezing of the orthogonal quadrature \( \op{x}_{\phi_-} = \op{x}_{\phi_+ + \frac{\pi}{2}} \).
This follows as \( r > 0 \) and \( \phi \in [0,\pi] \) and \( r \in \mathbb{R} \) and \( \phi \in [0,\pi/2] \) are equivalent parameterisations of the same squeezings---squeezing a quadrature \( \op{x}_{\phi} \) is equivalent to anti-squeezing the quadrature \( \op{x}_{\phi+\pi/2} \).

As \( B(Z_+) \) is a maximum and \( Z_- < 0 \) is outside the range of \( Z(\phi) \) at least one of \( \phi_{\pm} \) is a minimum of \( B(\phi) \).
We therefore find the global minimum of \( B(\phi) \) by finding the smaller of \( B(\phi_+) \) and \( B(\phi_-) \).
For \( Z(\phi_{\pm}) \) 
\begin{equation}
	Z(\phi_{\pm}) = 
	\left[ \left( 1+\frac{\omegat^2}{3} \right) \cosh 2r \pm \frac{\sqrt{9+3\omegat^2+\omegat^4}}{3} \sinh 2r \right],
\end{equation}
where we note that exchanging \( \phi_+ \to \phi_- \) is equivalent to \( r \to -r \).

For these squeezing angles (\( \phi_{\pm} \)) the bound (Eq.~\eqref{eq:app:qcrb_squeeze}) is
\begin{equation}
\begin{aligned}
	(\Delta\Lambda)^2 \geq
	\LambdaSQL^2 &
	\left(
		\left\{\thermvar^2 + \omegat \thermvar \LambdaR \left[ \left( 1+\frac{\omegat^2}{3} \right) \cosh 2r \pm \frac{\sqrt{9+3\omegat^2+\omegat^4}}{3} \sinh 2r \right] + \frac{\omegat^4}{12} \LambdaR^2 \right\}^2
		- 1
	\right) \\
	&\mkern32mu\times
	\Bigg(
		\frac{\omegat^4}{12}\left\{ \thermvar^2 
		+ \omegat \thermvar \LambdaR \left[ \left( 1+\frac{\omegat^2}{3} \right) \cosh 2r \pm \frac{\sqrt{9+3\omegat^2+\omegat^4}}{3} \sinh 2r \right]
		+ \frac{\omegat^4}{12} \LambdaR^2 \right\} \\
		&\mkern96mu +\frac{\omegat^4}{12}\left( 1-2\thermvar^2 \right) + \frac{\omegat^2}{2} \thermvar^2 \left[ \left( 1+\frac{\omegat^2}{3} \right) \cosh 2r \pm \frac{\sqrt{9+3\omegat^2+\omegat^4}}{3} \sinh 2r \right]^2
	\Bigg)^{-1},
\end{aligned}
\label{eq:app:qcrb_optimal_squeeze}
\end{equation}
which can be written as
\begin{equation}
	\frac{(a \pm b)^2 - 1}{c \pm d},
\end{equation}
where
\begin{align}
	a &= \thermvar^2 + \thermvar \LambdaR \omegat \left( 1 + \frac{\omegat^2}{3} \right) \cosh 2r + \frac{\omegat^4}{12} \LambdaR^2, \\
	b &= \thermvar \LambdaR \omegat \frac{\sqrt{9+3\omegat^2+\omegat^4}}{3} \sinh 2r, \\
	c &= \frac{\omegat^4}{12} \left[ 1 - \thermvar^2 + \thermvar \LambdaR \omegat \left( 1 + \frac{\omegat^2}{3} \right) \cosh 2r + \frac{\omegat^4}{12} \LambdaR^2 \right] +  \frac{\omegat^2}{2} \thermvar^2 \left[ \left( 1 + \frac{\omegat^2}{3} \right)^2 \cosh^2 2r + \left( 1 + \frac{\omegat^2}{3} + \frac{\omegat^4}{9} \right) \sinh^2 2r \right], \\
	d &= \thermvar \LambdaR \frac{\omegat^5}{12} \frac{\sqrt{9+3\omegat^2+\omegat^4}}{3} \sinh 2r + \thermvar^2 \omegat^2 \left( 1 + \frac{\omegat^2}{3} \right) \frac{\sqrt{9+3\omegat^2+\omegat^4}}{3} \cosh 2r \sinh 2r,
\end{align}
where we have \( a \), \( b \), \( c \), and \( d \) all positive as well as \( a > b+1 \) and \( c > d \).
The squeezing angle \( \phi_+ \) therefore offers a better precision for 
\begin{equation}
	c < d \left( \frac{a^2+b^2-1}{2ab} \right),
\end{equation}
which in this case is
\begin{equation}
	\begin{aligned}
	0 > 
	&\LambdaR \omegat \left[ - \thermvar^4 \left( 1 + 3\frac{\omegat^2}{4} + \frac{\omegat^4}{9} \right) + \frac{\omegat^2}{12} \left( 1 + \frac{\LambdaR^2\omegat^4}{12} \right)^2 + \thermvar^2 \frac{\omegat^2}{6} \left( 1 - \frac{\LambdaR^2\omegat^4}{12} \right) \right] \\
&+ \thermvar \left( 1 + \frac{\omegat^2}{3} \right) \left[ 1 - \thermvar^4 + \frac{\LambdaR^2\omegat^4}{6} (1-2\thermvar^2) + \left( \frac{\LambdaR^2\omegat^4}{12} \right)^2 \right] \cosh 2r
	- \frac{\omegat^3}{6} \thermvar^4 \LambdaR \cosh 4r.
	\end{aligned}
\end{equation}

\subsection{Homodyne detection}
The \ac{CRB} for homodyne measurement of the quadrature \( \op{x}\cos\theta + \op{p}\sin\theta \) is
\begin{equation}
\begin{aligned}
	(\Delta\Lambda)^2 \geq 
	2\LambdaSQL^2 \Bigg[
		\LambdaR + \thermvar 
		\Bigg(&
			\frac{
				\left[ 1 + \omegat^2 \right]\cos^2\theta + \omegat \sin 2\theta + \sin^2\theta
			}{
				\frac{1}{3}\omegat^3\cos^2\theta + \frac{1}{2}\omegat^2 \sin 2\theta + \omegat\sin^2\theta
			} \cosh 2r \\
			&+
			\frac{
				\left\{ \left[ 1-\omegat^2 \right]\cos^2\theta - \omegat \sin 2\theta - \sin^2\theta \right\}\cos 2\phi
				+
				\left\{ 2\omegat\cos^2 \theta + \sin 2\theta \right\}\sin 2\phi
			}{
				\frac{1}{3}\omegat^3\cos^2\theta + \frac{1}{2}\omegat^2 \sin 2\theta + \omegat\sin^2\theta
			} \sinh 2r
		\Bigg)
	\Bigg]^2.
\end{aligned}
\label{eq:app:crb_homodyne_squeezed}
\end{equation}

\subsubsection{Optimal squeezing}
The bound is minimised with respect to the squeezing angle \( \phi \) by minimising the coefficient of \( \sinh 2r \)
\begin{equation}
	\left[ \left( 1-\omegat^2 \right)\cos^2\theta - \omegat \sin 2\theta - \sin^2\theta \right]\cos 2\phi
	+
	\left[ 2\omegat\cos^2 \theta + \sin 2\theta \right]\sin 2\phi,
\end{equation}
which has minima
\begin{equation}
	\phi = - \arctan \left( \frac{1}{\omegat + \tan \theta} \right),
\end{equation}
for which squeezing angle the \ac{CRB} becomes
\begin{equation}
	(\Delta\Lambda)^2 \geq 
	2\LambdaSQL^2 \left[
		\LambdaR + e^{-2r} \thermvar 
		\left(
			\frac{
				\left[ 1 + \omegat^2 \right]\cos^2\theta + \omegat \sin 2\theta + \sin^2\theta
			}{
				\frac{1}{3}\omegat^3\cos^2\theta + \frac{1}{2}\omegat^2 \sin 2\theta + \omegat\sin^2\theta
			}
	\right)		
\right]^2.
\label{eq:app:crb_homodyne_optimally_squeezed}
\end{equation}

The optimal homodyne detection can then be recognised as the angle \( \theta \)
\begin{equation}
	\theta = - \arctan \left( \frac{3+2\omegat^2+\sqrt{9+3\omegat^2+\omegat^4}}{3\omegat} \right),
	\label{eq:app:optimal_homodyne}
\end{equation}
when this homodyne angle is used the optimal squeezing angle is
\begin{equation}
	\varphi = \arctan \left( \frac{3\omegat}{3-\omegat^2 + \sqrt{9+3\omegat^2+\omegat^4}} \right).
\end{equation}

\subsubsection{Position and Momentum squeezing}
Squeezing of position and momentum can be evaluated with \( \phi = 0 \), with \( r > 0 \) corresponding to squeezing of momentum while \( r < 0 \) is a squeezing \( |r| \) of position.
For \( \phi = 0 \) the \ac{CRB} (Eq.~\eqref{eq:app:crb_homodyne_squeezed}) becomes
\begin{equation}
	(\Delta\Lambda)^2 \geq 
	2\LambdaSQL^2 \Bigg[
		\LambdaR + \thermvar 
		\left(
			\frac{e^{2r} \cos^2 \theta + e^{-2r}\left(\omegat^2\cos^2 \theta+ \omegat \sin 2\theta + \sin^2 \theta\right)
			}{
				\frac{1}{3}\omegat^3\cos^2\theta + \frac{1}{2}\omegat^2 \sin 2\theta + \omegat\sin^2\theta
			}
		\right)
	\Bigg]^2.
\end{equation}
The optimal homodyne quadrature is then
\begin{equation}
	\theta = - \arctan \left( \frac{3e^{4r} + 2\omegat^2 + \sqrt{9e^{8r}+3e^{4r}\omegat^2+\omegat^4}}{3\omegat} \right),
\end{equation}
which gives a precision
\begin{equation}
	(\Delta\Lambda)^2 \geq
	2\LambdaSQL^2 \left[ 
	\LambdaR + \thermvar
	\left( \frac{2(3e^{2r} + e^{-2r}\omegat^2 - \sqrt{9e^{4r} + 3\omegat^2 + e^{-4r}\omegat^4})}{\omegat^3} \right)
	\right],
\end{equation}
where squeezing of position (\( r < 0 \)) is beneficial for \( \omegat < \sqrt{3} \) while squeezing of momentum (anti-squeezing of position, \( r > 0 \)) is beneficial for \( \omegat > \sqrt{3} \).

\section{Heterodyne detection}
\label{app:heterodyne}
Heterodyne detection is the projection onto the overcomplete basis of Gaussian states which amounts to sampling from the Husimi Q-function~\cite{shapiro_phase_1984,leonhardt_measuring_1995}.
The Q-function can be extracted from the Wigner function as~\cite{ferraro_gaussian_2005}
\begin{equation}
	Q(x,p)=\frac{1}{\pi} \int\mathrm{d}x'\mkern-1mu\mathrm{d}p'\mkern-1mu W(x',p') \exp \left[ -(x-x')^2-(p-p')^2 \right].
\end{equation}
which is a convolution and so for a Gaussian Wigner function with moments \( \vec{d} \) and \( \sigma \) the Q function will be Gaussian with moments \( \vec{d} \) and \( \sigma + \mathbb{1} \)~\cite[Chap.~5]{serafini_quantum_2017}.

The mean of the distribution again contains no parameter dependence and so Eq.~\eqref{eq:gaussian_cfi} can also be applied here.
The covariances from heterodyne detection are
\begin{equation}
	\Sigma(\omegat) = 
	\begin{pmatrix}
		1 + \Sigma_{xx} + 2\omegat\Sigma_{xp} + \omegat^2 \Sigma_{pp} + \frac{1}{3} \LambdaR \omegat^3 &
		\Sigma_{xp} + \omegat\Sigma_{pp} + \frac{1}{2} \LambdaR \omegat^2 \\
		\Sigma_{xp} + \omegat\Sigma_{pp} + \frac{1}{2} \LambdaR \omegat^2 &
		1 + \Sigma_{pp} + \LambdaR \omegat
	\end{pmatrix},
	\label{eq:heterodyne_covariance}
\end{equation}
giving a \ac{CRB} of
\begin{equation}
	(\Delta\Lambda)^2 \geq
	\frac{12\LambdaSQL^2|\Sigma(\omegat)|^2}{\omegat^4 |\Sigma(\omegat)| + 6 \omegat^2 \left( 1 + \Sigma_{xx} + \omegat \Sigma_{xp} + \frac{\omegat^2}{3}\Sigma_{pp} \right)^2 + 2 \omegat^4 \left[ 1 + \Sigma_{xx} - \Sigma_{pp} - \Sigma_{xx}\Sigma_{pp} + \Sigma_{xp}^2 + \frac{\omegat^2}{3}\left( 1-\Sigma_{pp} \right) \right] },
	\label{eq:app:heterodyne_crb}
\end{equation}
where \( |\Sigma| \) is the determinant, and \( \Sigma_{xx} \), \( \Sigma_{xp} \), and \( \Sigma_{pp} \) are the initial variances and covariance of the position and momentum operators.
Without mechanical squeezing (\( r=0 \)) this is
\begin{equation}
	(\Delta\Lambda)^2 \geq
	\frac{
		6 \LambdaSQL^2 \left[ (1+\thermvar)^2 + \thermvar \omegat^2 + \LambdaR(1+\thermvar)\omegat \left(1+\frac{\omegat^2}{3}\right) + \frac{\omegat^4 }{12} \LambdaR^2 \right]^2
	}{
		\frac{\omegat^2}{3} \left[ (1+\thermvar^2)(9+3\omegat^2+\omegat^4) + \thermvar (18 + 6\omegat^2-\omegat^4) \right] + \frac{\omegat^4}{2} \left[ (1+\thermvar)^2 + \thermvar \omegat^2 + \LambdaR(1+\thermvar)\omegat \left(1+\frac{\omegat^2}{3}\right) + \frac{\omegat^4}{12} \LambdaR^2 \right]
	}.
\end{equation}

\section{Optimal measurement}
\label{app:sld_projectors}
For a Gaussian system the \ac{SLD} is a hermitian operator, quadratic in the quadrature operators~\cite{monras_phase_2013,serafini_quantum_2017}.
Any such hermitian operator, quadratic in the quadrature operators, can be transformed through some squeezing and displacement to an operator diagonal in the Fock basis~\cite{monras_phase_2013,serafini_quantum_2017}.

The \ac{SLD} is primarily defined through identification of \( L^{(2)} \) which is given by~\cite{monras_phase_2013,serafini_quantum_2017}
\begin{equation}
	\sigma L^{(2)} \sigma + \Omega L^{(2)} \Omega = \partial \sigma,
	\label{eq:app:ltwo}
\end{equation}
which for a state with constant zero displacements \( \vec{d} = 0 \) then gives the \ac{SLD}~\cite{monras_phase_2013,serafini_quantum_2017}
\begin{equation}
	L_{\rho_{\Lambda}} = \begin{pmatrix} \op{x} & \op{p} \end{pmatrix} L^{(2)} \begin{pmatrix} \op{x} \\ \op{p} \end{pmatrix} - \frac{1}{2}\Trace{L^{(2)}\sigma}.
\end{equation}
The covariance matrix which we wish to solve for Eq.~\eqref{eq:app:ltwo} is Eq.~\eqref{eq:cov}, which gives \( L^{(2)} \) as
\begin{equation}
		L^{(2)} = 
	\frac{1}{\LambdaSQL(|\sigma(\omegat)|^2-1)}
	\begin{pmatrix}
		l^{(2)}_{xx} & l^{(2)}_{xp} \\
		l^{(2)}_{xp} & l^{(2)}_{pp}
	\end{pmatrix},
\end{equation}
where
\begin{align}
	l^{(2)}_{xx} &= \omegat+\omegat \sigma_{xp}(\omegat)^2 - \omegat^2 \sigma_{xp}(\omegat)\sigma_{pp}(\omegat) + \omegat^3 \sigma_{pp}(\omegat)^2, \\
	l^{(2)}_{xp} &= -\omegat \sigma_{xx}(\omegat) \sigma_{xp}(\omegat) + \frac{\omegat^2}{2} \left( \sigma_{xx}(\omegat) \sigma_{pp}(\omegat) + \sigma_{xp}(\omegat)^2-1 \right) - \frac{\omegat^3}{3} \sigma_{xp}(\omegat)\sigma_{pp}(\omegat), \\
	l^{(2)}_{pp} &= \omegat \sigma_{xx}(\omegat)^2 - \omegat^2 \sigma_{xx}(\omegat)\sigma_{xp}(\omegat) + \frac{\omegat^3}{3}(1+\sigma_{xp}(\omegat)^2).
\end{align}
Then \( L^{(2)} \) has eigenvalues
\begin{equation}
	\alpha \pm \sqrt{\alpha^2 - \omegat^2 \left( \sigma_{xx}(\omegat) - \omegat \sigma_{xp}(\omegat) + \frac{\omegat^2}{3} \sigma_{pp}(\omegat) \right)^2 - \frac{\omegat^4}{12}\left( |\sigma(\omegat)|-1 \right)^2},
\end{equation}
with
\begin{equation}
	\alpha(\omegat) = \frac{\omegat}{2}\left[ 1 + \sigma_{xx}(\omegat)^2 - \sigma_{xp}(\omegat)(\sigma_{xx}(\omegat)+\sigma_{pp}(\omegat)) \omegat + \frac{\omegat^2}{3}(1+\sigma_{pp}(\omegat)^2) + \sigma_{xp}(\omegat)^2 \left( 1 + \frac{\omegat^2}{3} \right) \right].
\end{equation}
In order for phonon-number resolving detection to become optimal we then seek the symplectic transformation which gives the Williamson normal form of \( L^{(2)} \).
For a single-mode system this can be recognised by first diagonalising \( L^{(2)} \) with a phase shift
\( 
	\begin{pmatrix} \cos\psi & \sin\psi \\ -\sin\psi & \cos\psi \end{pmatrix}
\),
followed by a squeezing \( \diag (e^{z} , e^{-z}) \).
The phase shift diagonalises \( L^{(2)} \), which has eigenvalues \( D_1 \) and \( D_2 \).
The symplectic eigenvalue of \( L^{(2)} \) is then \( \sqrt{D_1D_2} \) and so the squeezing \( z \) required to bring \( L^{(2)} \) into its normal form is \( e^{2z} = e^{\frac{1}{2}|\ln D_1 - \ln D_2|} \).

Thus the required squeezing is
\begin{equation}
	e^{2z} = \sqrt{\frac{
			1 + \sqrt{1 - \frac{1}{\alpha^2} \left[ \omegat^2 \left( \sigma_{xx}(\omegat) - \omegat \sigma_{xp}(\omegat) + \frac{\omegat^2}{3} \sigma_{pp}(\omegat) \right)^2 + \frac{\omegat^4}{12}\left( |\sigma(\omegat)|-1 \right)^2 \right]}
}{
			1 - \sqrt{1 - \frac{1}{\alpha^2} \left[ \omegat^2 \left( \sigma_{xx}(\omegat) - \omegat \sigma_{xp}(\omegat) + \frac{\omegat^2}{3} \sigma_{pp}(\omegat) \right)^2 + \frac{\omegat^4}{12}\left( |\sigma(\omegat)|-1 \right)^2 \right]}
	}}.
\end{equation}

\section{Optimality of detection schemes}
\label{app:relative_precisions}
Our bounds cover a range of settings with \( \LambdaSQL^2 \) pre-factoring the bounds and their ratios being a function of only \( \LambdaR \), \( \omegat \), \( \thermvar \), and squeezing \( re^{i\phi} \) (with parameters such as homodyne angle \( \theta \) representing different measurement choices rather than properties of the system).
This allows comparison of our bounds in terms of these parameters alone, perhaps the simplest case being where we assume trapping allows us to take \( \thermvar = 1 \) and that no external squeezing is applied

\subsection{Homodyne}
\begin{figure}[htbp]
	\centering
		\includegraphics{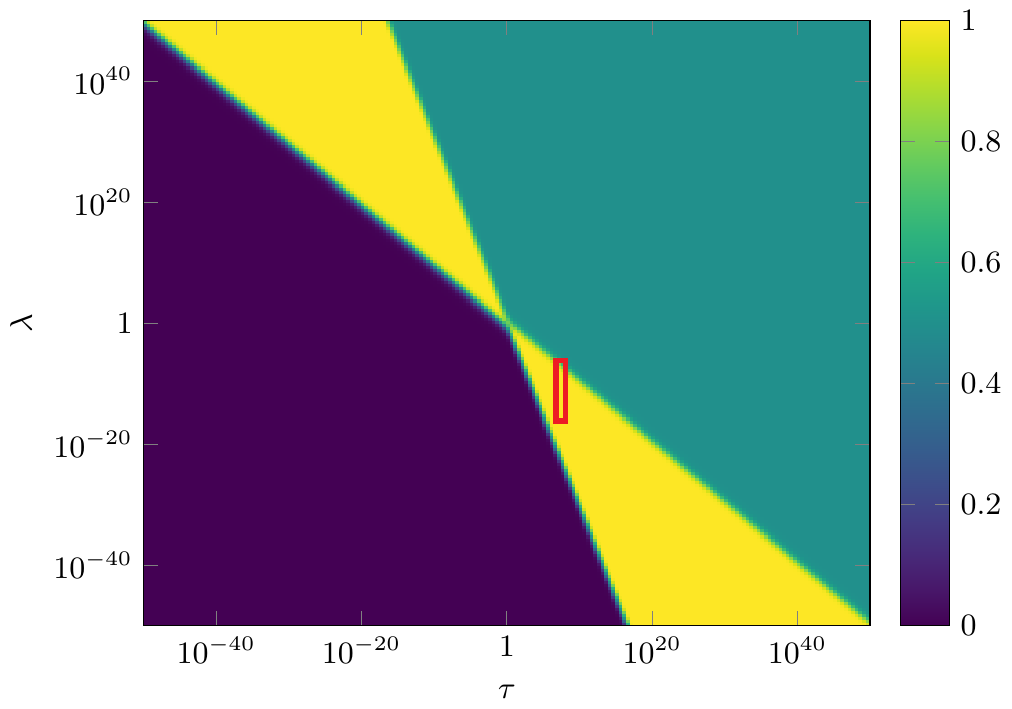}
		\caption{Ratio of quantum Fisher information against classical Fisher information for optimal homodyne quadrature (\( F(\Lambda;\theta_{\textrm{opt}})/H(\Lambda) \)), plotted for \( \thermvar = 1 \) and \( r = 0 \).
		The red rectangle is representative of the MAQRO parameter regime}
	\label{fig:homodyne_optimality}
\end{figure}

For \( \thermvar = 1 \) and \( r = 0 \) we can easily compare the \ac{QCRB} with the optimal homodyne \ac{CRB} numerically across the \( \LambdaR \) and \( \omegat \) variables in Fig.~\ref{fig:homodyne_optimality}.
The analytic form of the ratio is
\begin{equation}
R = \frac{ \omegat^4 \left\{ \left[ \LambdaR \omegat \left( 1 + \frac{\omegat^2}{3} + \frac{\omegat^3}{12}\LambdaR \right) + 1 \right]^2 - 1 \right\}
}{
	72 \left( 1 + \frac{\omegat^2}{3} + \frac{\omegat^3}{6} \LambdaR - \frac{\sqrt{9+3\omegat^2+\omegat^4}}{3}\right)^2\left[ \left( 1 + \frac{\omegat^2}{3} + \frac{\omegat^3}{12} \LambdaR \right)^2 - \frac{1}{2} \left( 1 + \frac{\omegat^2}{3} \right) \left( 1 + \frac{\omegat^2}{3} + \frac{\omegat^3}{6} \LambdaR \right) \right]
}.
\label{eq:app:hom_q}
\end{equation}

\subsection{Heterodyne}
\begin{figure}[htbp]
	\centering
	\includegraphics{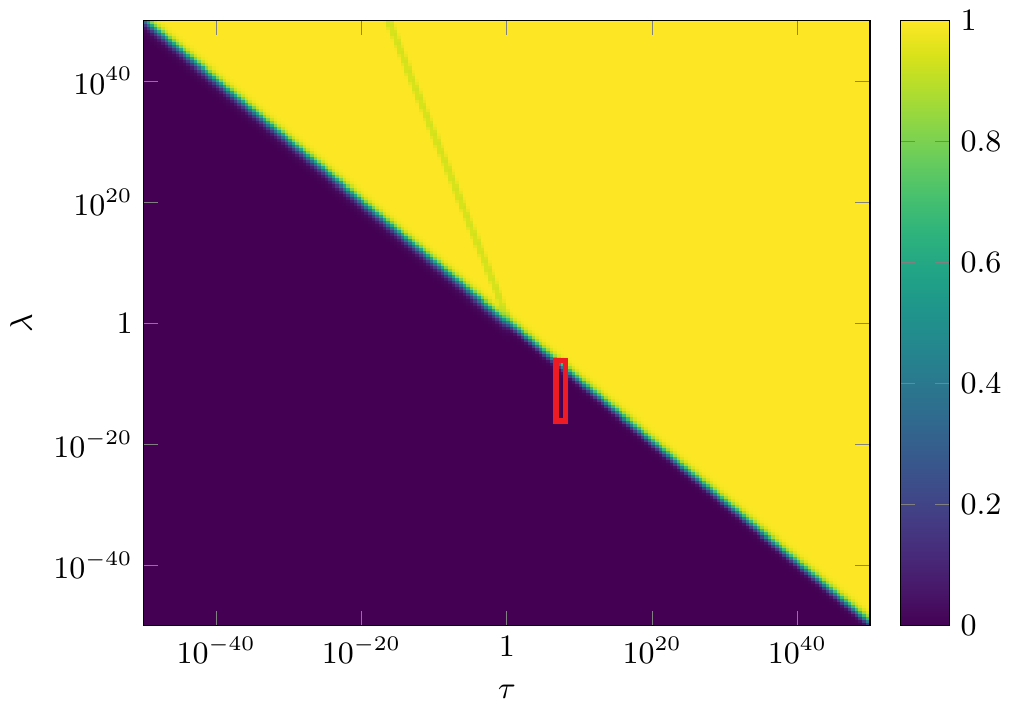}
	\caption{Ratio of quantum Fisher information against classical Fisher information for heterodyne detection (\( F(\Lambda)/H(\Lambda) \)), plotted for \( \thermvar = 1 \) and \( r = 0 \).
		The red rectangle is representative of the MAQRO parameter regime}
	\label{fig:heterodyne_optimality}
\end{figure}

For \( \thermvar = 1 \) and \( r = 0 \) we can easily compare the \ac{QCRB} with the heterodyne \ac{CRB} numerically across the \( \LambdaR \) and \( \omegat \) variables in Fig.~\ref{fig:heterodyne_optimality}.
The analytic form of the ratio is
\begin{equation}
	R=
	\frac{\left\{\left[\LambdaR \omegat \left(1+\frac{\omegat^2}{3}+\LambdaR\frac{\omegat^3}{12}\right)+1\right]^2-1\right\} \left[\left(1+\frac{\omegat^2}{3}+\LambdaR\frac{\omegat^3}{12}\right)^2+\left(1+\frac{\omegat^2}{6}\right)^2\right]
	}{
	16 (1 + \frac{\omegat}{2}\LambdaR)^2 \left(1+\frac{\omegat^2}{4}+\frac{\LambdaR \omegat^3}{24}\right)^2 \left[\left(1+\frac{\omegat^2}{3}+\frac{\LambdaR \omegat^3}{12}\right)^2-\frac{1}{2} \left(1+\frac{\omegat^2}{3}\right) \left(1+\frac{\omegat^2}{3}+\frac{\LambdaR \omegat^3}{6}\right)\right]
}.
\label{eq:app:het_q}
\end{equation}

\subsection{Homodyne and Heterodyne}
\begin{figure}[htbp]
	\centering
	\includegraphics{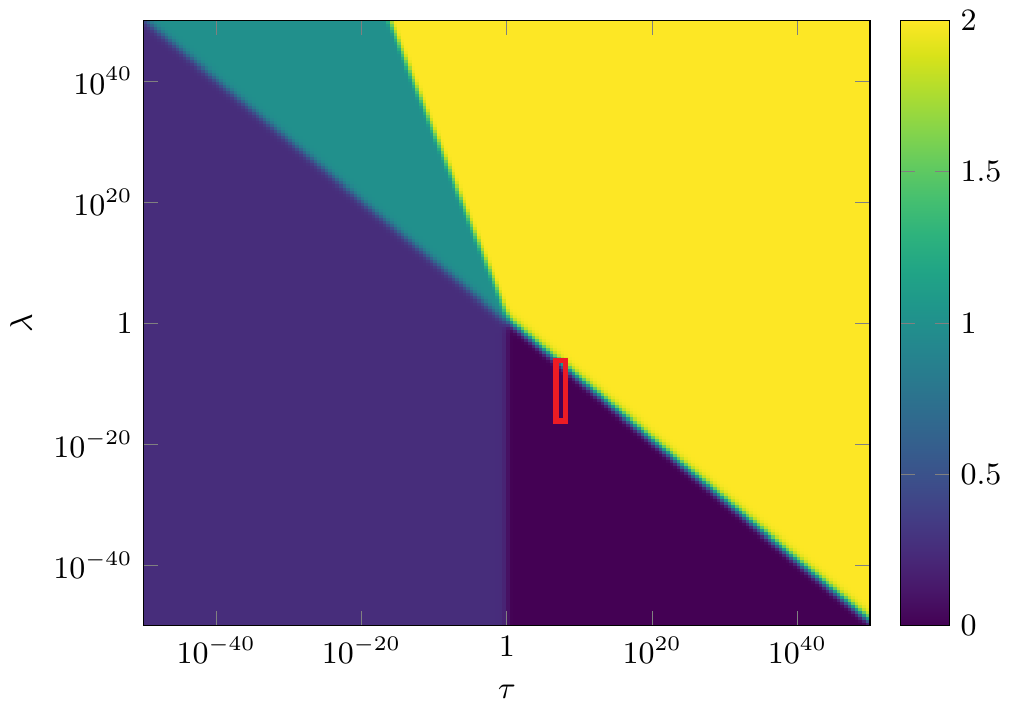}
	\caption{Ratio of classical Fisher information for heterodyne detection against classical Fisher information for homodyne detection of the optimal quadrature, plotted for \( \thermvar = 1 \) and \( r = 0 \).
		The red rectangle is representative of the MAQRO parameter regime}
	\label{fig:homodyne_heterodyne_optimality}
\end{figure}

In the same \( \thermvar = 1 \) and \( r = 0 \) case we can compare the optimal homodyne \ac{CRB} against the heterodyne \ac{CRB} numerically across the \( \LambdaR \) and \( \omegat \) variables in Fig.~\ref{fig:homodyne_heterodyne_optimality}.
This demonstrates no more than a factor of two advantage for heterodyne in the \( \omegat \gg 1 \) and \( \LambdaR \gg 1 \), while in the \( \LambdaR \ll 1 \) regime homodyne has a near unbounded advantage.

The analytic form of the ratio (which can be seen from Eqs.~\eqref{eq:app:hom_q} and~\eqref{eq:app:het_q}) is
\begin{equation}
	R=
	\frac{
		9 \left[\left(1+\frac{\omegat^2}{3}+\LambdaR\frac{\omegat^3}{12}\right)^2+\left(1+\frac{\omegat^2}{6}\right)^2\right]
	\left( 1 + \frac{\omegat^2}{3} + \frac{\omegat^3}{6} \LambdaR - \frac{\sqrt{9+3\omegat^2+\omegat^4}}{3}\right)^2
	}{
	2 \omegat^4(1 + \frac{\omegat}{2}\LambdaR)^2 \left(1+\frac{\omegat^2}{4}+\frac{\LambdaR \omegat^3}{24}\right)^2 
}.
	\label{eq:app:hom_het}
\end{equation}

\section{Tests of Continuous Spontaneous Localisation}
\label{app:csl}
\begin{figure}[htbp]
\centering
\includegraphics{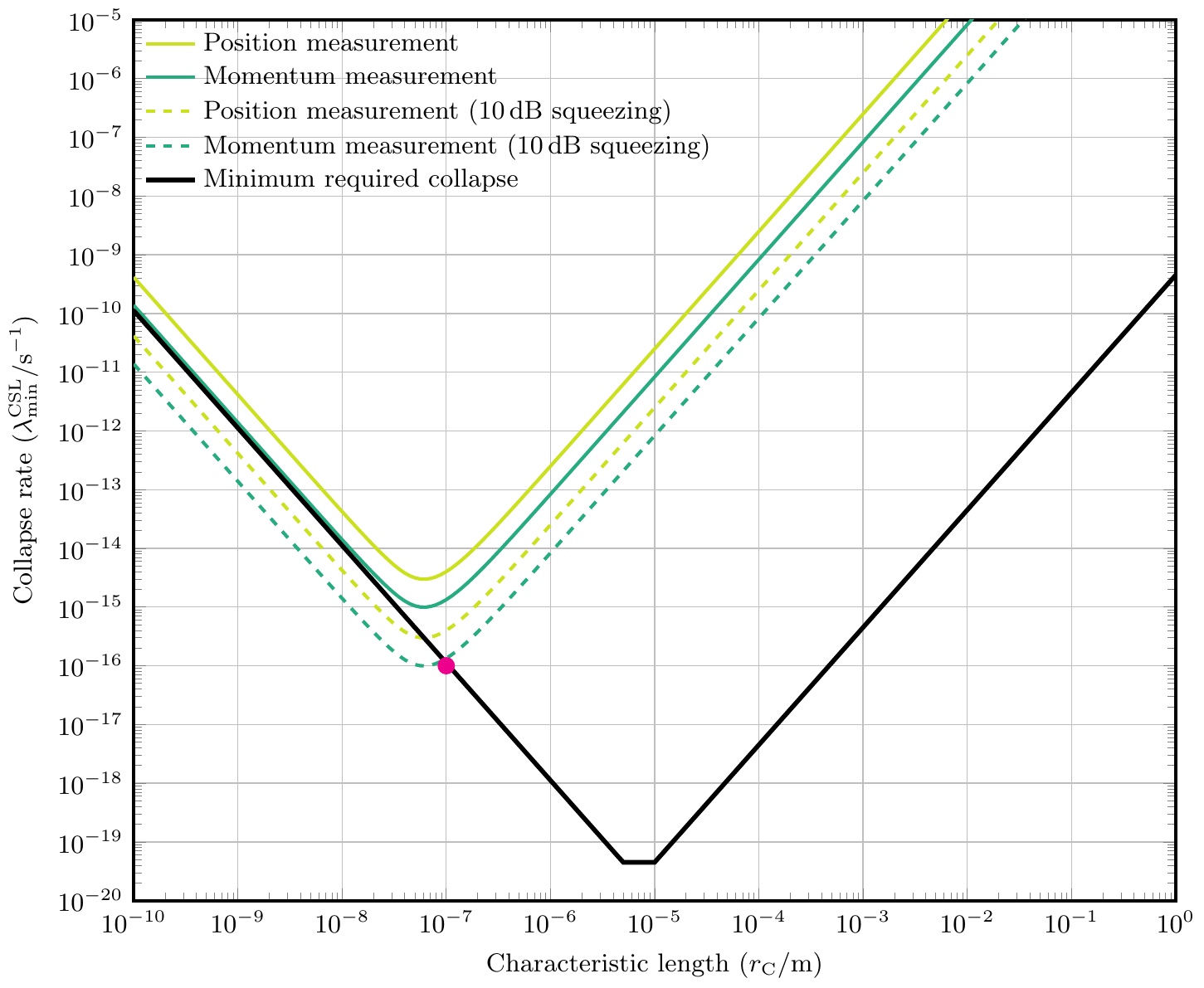}
\caption{
	Bounds plotted for a \( \rS = \SI{100}{\nm} \) sphere of mass \SI{5.5e9}{\amu} with values otherwise as Tab.~\ref{tab:maqro_values} in the main text.
	The minimum required collapse rate given is based on the criteria of Ref.~\cite{toros_colored_2017}.
	The magenta dot represents the values originally proposed by \citet{ghirardi_unified_1986}.
}
\label{app:fig:csl_conventional}
\end{figure}

For MAQRO the minimum resolvable \( \lambdaCSL \) for position and momentum can be seen in Fig.~\ref{app:fig:csl_conventional}, plotted for a \( \rS = \SI{100}{\nm} \) sphere of mass \SI{5.5e9}{\amu} with values otherwise as Tab.~\ref{tab:maqro_values} in the main text, where the black line is based on the minimum required \ac{CSL} strenght proposed in \citet{toros_colored_2017}.
This plot shows the potential improvments, with MAQRO already competitive in \SIrange[range-units=single]{1e-8}{1e-5}{\m}, squeezing allows a test down to the lower bound for \( \rC < \SI{1e-7}{\m} \) and significant improvement on reported results up to \( \rC = \SI{1e-5}{\m} \).

\begin{figure}[htbp]
\centering
\includegraphics{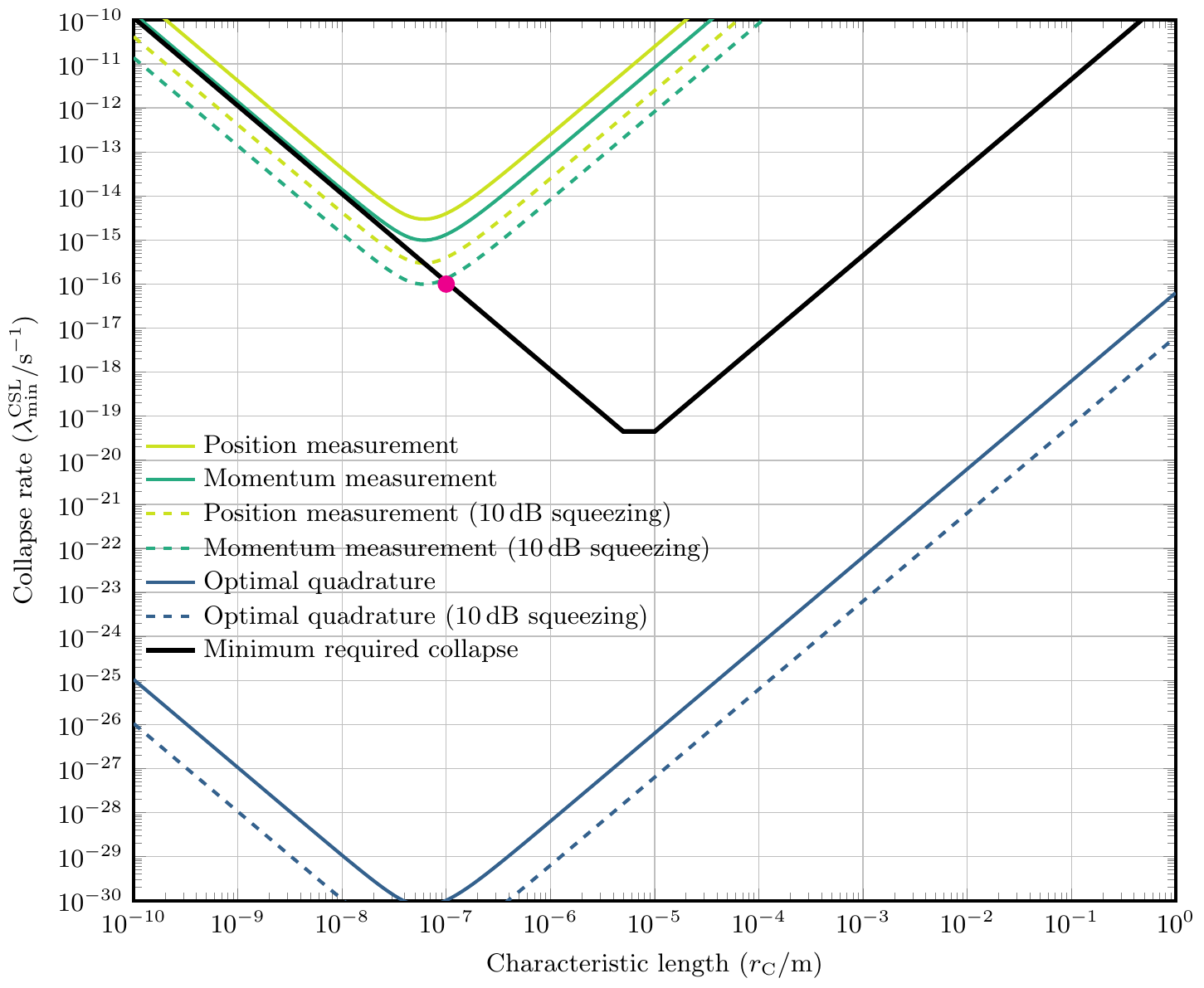}
\caption{
	Bounds plotted for a \( \rS = \SI{100}{\nm} \) sphere of mass \SI{5.5e9}{\amu} with values otherwise as Tab.~\ref{tab:maqro_values} in the main text.
	The minimum required collapse rate given is based on the criteria of Ref.~\cite{toros_colored_2017}.
	The magenta dot represents the values originally proposed by \citet{ghirardi_unified_1986}.
}
\label{app:fig:csl_optimal}
\end{figure}

This is plotted in Fig.~\ref{app:fig:csl_optimal}, plotted again for a \( \rS = \SI{100}{\nm} \) sphere of mass \SI{5.5e9}{\amu} with values otherwise as Tab.~\ref{tab:maqro_values} in the main text.
As might be guessed from the significant gap in Fig.~\ref{fig:maqro_precision} of the main text the optimal quadrature allows for a categorical test of \ac{CSL}. 
This bound can be reduced through squeezing and the fundamental limit given by the \ac{QCRB} will further allow a superior precision through a saturating measurement.
Such improvements however offer little significance, as the \ac{QCRB} will give a lower bound no less than that of the optimal quadrature.

\end{document}